\documentclass[12pt]{iopart}
\usepackage{bm}
\usepackage{graphicx}
\usepackage{dsfont}

\usepackage{color}
\begin{document}

\title{Effective geometric phases and topological transitions in SO(3) and SU(2) rotations}

\author{Henri Saarikoski,$^1$  Jos\'e Pablo Baltan\'as,$^2$ J. Enrique V\'azquez-Lozano,$^2$  Junsaku Nitta,$^3$ and Diego Frustaglia$^2$}
\address{$^1$RIKEN Center for Emergent Matter Science (CEMS), Saitama 351-0198, Japan}
\ead{e-mail: henri.saarikoski@riken.jp}
\address{$^2$Departamento de F\'isica Aplicada II, Universidad de Sevilla, E-41012 Sevilla, Spain}
\ead{e-mail: frustaglia@us.es}
\address{$^3$Department of Materials Science, Tohoku University, Sendai 980-8579, Japan}
\begin {abstract} {We address the development of geometric phases in
classical and quantum magnetic moments (spin-1/2) precessing in an
external magnetic field.  We show that nonadiabatic dynamics lead
to a topological phase transition determined by a change in the
driving field topology.  The transition is associated with an
effective geometric phase which is identified from the paths of
the magnetic moments in a spherical geometry.  The topological
transition presents close similarities between SO(3) and SU(2)
cases but features differences in, {\em e.g.}, { the
adiabatic limits of the geometric phases, being $2\pi$ and
$\pi$ in the classical and the quantum case, respectively}.  We
discuss possible experiments where the effective geometric phase
would be observable.}
\end{abstract}

\date{\today}


\maketitle
\section{Introduction}
An adiabatic cyclic process involves slowly changing periodic parameters driving a physical system.
In the absence of transitions the system returns to its initial state after the cycle. { However, the quantum state may acquire a phase factor in this process} 
that can be written in general as the sum of a dynamical phase that depends on the Hamiltonian and an extra term of geometrical origin.
The latter is known as the Berry phase and can be calculated
from the geometric properties of the path described by the driving parameters in the parameter space of the Hamiltonian~\cite{berry}.
Nonadiabatic cyclic processes may also give rise to geometric phases that are smaller than adiabatic Berry phases
but with similar geometric characteristics~\cite{aharonov,bohm}.
Geometric phases can be observed experimentally
via interference of waves acquiring different phase components, {\em e.g.}, in the Aharonov--Bohm effect~\cite{chambers,osakabe}
and its electromagnetic dual, the Aharonov--Casher effect~\cite{cimmino,nagasawa1,nittameijer,frustagliarichter,richter,nagasawa}.

{ Phases of geometrical origin} were anticipated in Pancharatnam's work on interference of polarized light~\cite{pancharatnam}
and  they also have { been found} in classical mechanics~\cite{hannay}.
A famous example is parallel transport of vectors on a curved surface~\cite{bgoss}
where the geometric phase is associated with the rotation of a transported coordinate system. { Geometric phases and related topological effects are also ubiquitous in molecular physics. For example, they play a crucial role when the Born-Oppenheimer approximation breaks down near a conical intersection of two potential energy surfaces. In these cases, the electronic part of the wave function may acquire a phase shift when the nuclei traverse a closed path (see, \emph{e.g.} Refs.~\cite{mead,baer,juanesmarcos,joubertdoriol}). Such geometric phase effects lead to significant energy shifts~\cite{kendrick}, changes in reaction rates~\cite{kendrick2}, and localization of eigenstates~\cite{ryabinkin}}.

{ As pointed out above}, geometric phases depend on the geometric properties of { the path of} the system in the parameter space of the Hamiltonian.
Hence, they can be controlled via deformation of { this path}.
Lyanda-Geller proposed that the geometric phase could be switched on and off in spin interference experiments by changing
the topology of the spin-guiding field from rotating to nonrotating~\cite{lyanda-geller} (see Fig.~\ref{fieldtopo}).
{ This can be done by} using a combination of rotating and homogeneous fields of different relative strength.
Spin transport calculations predict that such a transition in the field topology causes a distinct shift in the interference pattern
of the conductance in loop-shaped spin-interferometers~\cite{saarikoskiprb}.
However, due to complex nonadiabatic spin dynamics,
the topological transition is associated with an emergent \emph{effective} geometric phase~\cite{saarikoskiprb} and not with the bare geometric phase
that can be calculated from the solid angle that the spin subtends in a roundtrip around the circuit.

Recently, Berry phases have also been extracted using
interferometry in superconducting qubit systems~\cite{leek}. Indeed, a
topological transition in such spin-${1 \over 2}$ superconducting
qubit systems has been observed by measuring the Chern number, a topological invariant that counts the number of degenerate energy
eigenvalues inside a closed manifold along which the parameters of
the Hamiltonian are swept~\cite{schroer2014,roushan}. Geometric phases have been studied
also in the context of nuclear magnetic resonance spectroscopy (NMR)~\cite{suter}
for which a general treatment was developed by Bloch and Siegert~\cite{blochsiegert}.
NMR beyond a perturbative regime has been considered in Refs.~\cite{shirley,swain}. In the experiments discussed above, 
the rotating field is usually assumed to be perpendicular to the homogeneous magnetic field~\cite{suter,bohm}.
In contrast, here we change the topology of the field 
using a coplanar arrangement of the homogeneous and rotating field components with arbitrary field strengths.
The changing field topology is then associated with nonadiabatic dynamics when a degeneracy is
placed along the path of the cycle, leading to emergent effective geometric phases undergoing a smooth but distinct transition.

Here we study the nonadiabatic dynamics of a classical and a quantum mechanical (spin)
magnetic moment precessing in a time-dependent magnetic
field with varying topology due to the combined action of coplanar, homogeneous and rotating, field components.
U(1) phases do not play any role in the classical dynamics (in contrast to the quantum case)
but the Lie algebras of the group SU(2) of spin rotations and SO(3) of classical rotations are isomorphic
leading to analogous phenomena. Cina demonstrated in a study on spin and classical magnetic moment
precessions that {\em adiabatic}  Berry phase factors have a geometric interpretation in terms of
rotations~\cite{cina}. Moreover, SO(3) has been found to be connected with adiabatic geometric phases
in qubit systems~\cite{milman}.  Therefore we expect analogies to emerge also in the nonadiabatic regime and 
we anticipate comparable topological effects in both cases.
Indeed, we show here that precessions of a classical magnetic moment give rise to effective nonadiabatic geometric and dynamic phase %
angles with an associated topological transition, showing a clear correspondence with those found for spin rotations.
The results also provide insights into the origin of the effective geometric phases in quantum systems~\cite{saarikoskiprb}.
We demonstrate that, in the classical case,
the effective geometric phase is associated with the dynamics of
precessions of the classical magnetic moment in the $\mathrm{S}^2$
sphere, much like it was found to be related to the windings of the
spin magnetic moment on the Bloch sphere in the quantum case~\cite{saarikoskiprb}.
In the adiabatic limit we recover a Berry phase equal to $2\pi$ for classical rotations which is in contrast to the Berry phase of $\pi$
for spin rotations.  We also { suggest} experimental setups that would demonstrate the discussed effects.

\section{Definitions and models}

\label{sec:methods}
\begin{figure}
\includegraphics[width=\columnwidth]{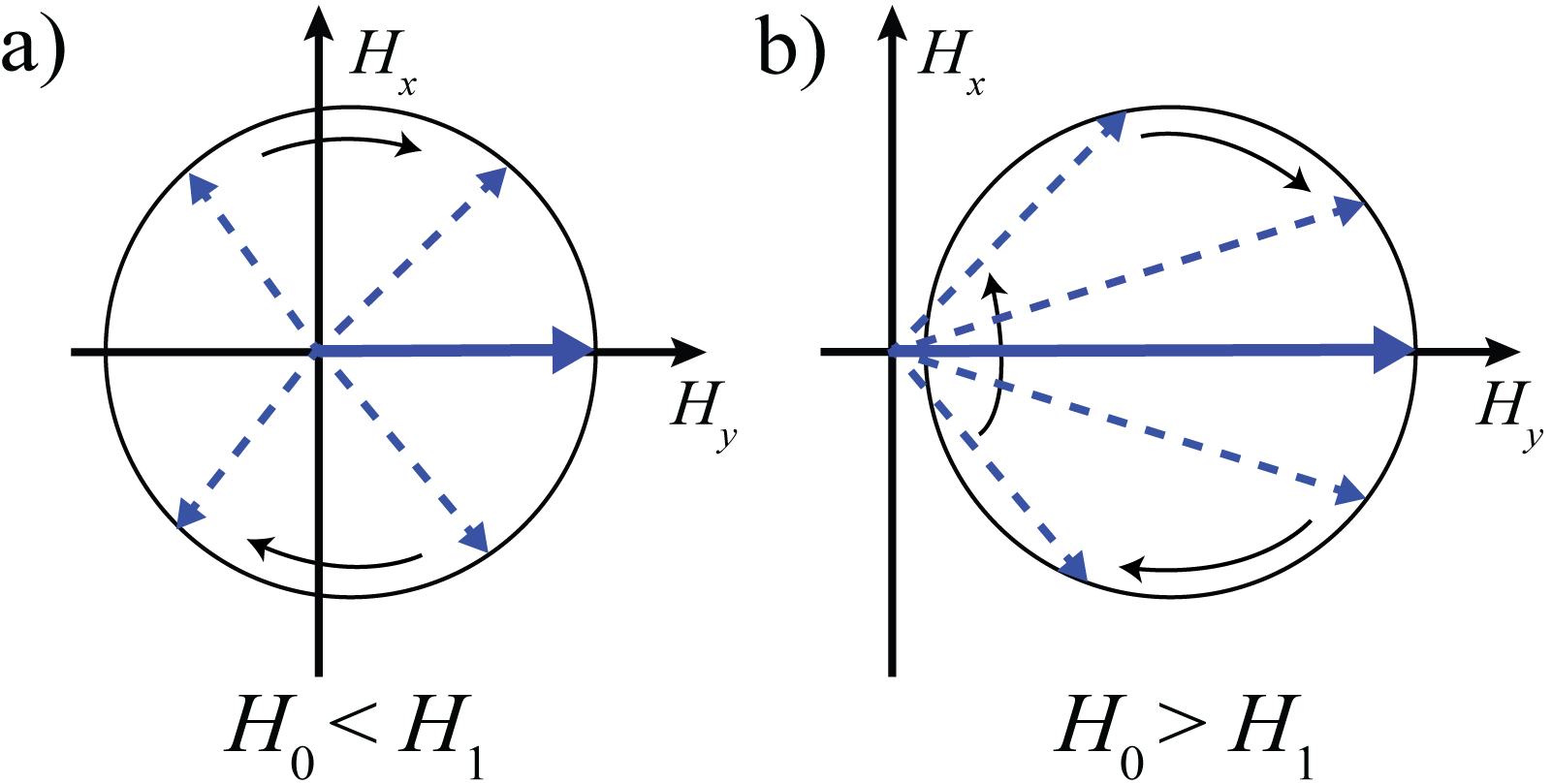}
\caption{Magnetic field ${\bf H}=2\pi(H_0{\bf e}_y+H_1(\sin\theta\,{\bf e}_x+\cos\theta\,{\bf e}_y))$
changes topology from rotating (a) to nonrotating (b) as $H_0$ becomes larger than $H_1$.
\label{fieldtopo}}
\end{figure}

We study the dynamics of a classical magnetic moment and a spin-$1 \over 2$ system in a composite magnetic field 
\begin{equation}
{\bf H}(t)={\bf H}_0+{\bf H}_1(t),
\label{magfield}
\end{equation}
where ${\bf H}_0= 2\pi H_0 \ {\bf e}_y$ is an homogeneous field and ${\bf H}_1(t)= 2\pi H_1 (\sin\omega t \ {\bf e}_x + \cos\omega t \ {\bf e}_y)$ is a time-dependent one
with $\omega$ the angular frequency, ${\bf e}_{x,y,z}$ the unit vectors along $x$, $y$ and $z$ axes, respectively, and $H_0$ and $H_1$ the strengths of the corresponding field components.
The coplanar field texture (\ref{magfield}) undergoes a change in topology from nonrotating ($H_0 > H_1$) to rotating ($H_0 < H_1$)
when the strengths of the field components are equal, $H_{0}=H_{1}$, see Fig.~\ref{fieldtopo}.
%

The magnetic moment dynamics in the classical case is calculated numerically using the Landau-Lifshitz equation~\cite{landaulifshitz} (or Bloch equation~\cite{bloch}).
We assume in the following that the magnetic moment ${\bf M}=M_x{\bf e}_{x}+M_y{\bf e}_{y}+M_z{\bf e}_{z}$  has unit length.
The total effective magnetic field ${\bf H}(t)$ then exerts a torque on
the magnetic moment giving rise to precessions as described by the equation of motion
\begin{equation}
\frac{d{\bf M}(t)}{dt}=g{\bf H}(t)\times{\bf M}(t),
\label{eq:LL}
\end{equation}
where $g$ is the gyromagnetic ratio that scales the effective field strength.
It does not play another role in the calculations of phase components. Therefore we set it equal to one without loss of generality.

The total rotation matrix in a roundtrip $\theta=0\to 2\pi$, with $\theta= \omega t$, is then a matrix product of infinitesimal rotations {
\begin{equation}
R=\prod_{m=1}^{n}\left (\begin{array}{ccc}
1 & 0 & H_y(\theta_{m})d\theta\\
0 & 1 & -H_x(\theta_{m}) d\theta\\
-H_y(\theta_{m})d\theta & H_x(\theta_{m}) d\theta& 1\\ 
 \end{array}
\right )
\label{class_matrix}
\end{equation}
in the limit $n\rightarrow\infty$. Here, $d\theta=2\pi/n$ and $\theta_{m}=(m-1)d\theta$}. Unlike in magnetic resonance models, we focus on the acquired phases during a cycle.
In this case, by scaling time as $t'=\omega t$ and the strength of the
magnetic fields as ${\bf H}'={\bf H}/\omega$, the solution of Eq.~(\ref{eq:LL})
becomes independent of $\omega$. Therefore we set $\omega=1$ in the following discussion.

We set ${\bf M}(t=0)={\bf M}_i$ and obtain numerically the magnetic moment after a rotation $\theta=0\to 2\pi$, ${\bf M}_f$.
To this end, we { fix $n$ in Eq.~(\ref{class_matrix}) and normalize ${\bf M}$ to unit length after each step.
In numerical calculations, $n$ is of the order of
$10\,000$ and convergence of the results was conveniently checked}. From this SO(3) rotation,
we define the total phase 
acquired during the evolution as the angle between the initial and final magnetic moment orientation
\begin{equation}
\phi_{\rm tot}=\angle({\bf M}_i,{\bf M}_f)=\angle({\bf M}(\theta=0),{\bf M}(\theta=2\pi)).
\label{totphase}
\end{equation}
In addition, we define the bare dynamic phase $\phi_{\rm d}$ as the total rotation angle of the magnetic moment in the driving field.
This is equal to the path length that the tip of the magnetic moment vector sweeps on the unit sphere in a cycle
\begin{equation}
\phi_{\rm d}= \int |d{\bf M}(\theta)|.
\label{dynamicso3}
\end{equation}
Finally, the bare geometric phase $\phi_{\rm bg}$ is defined via
\begin{equation}
\phi_{\rm tot}=\phi_{\rm d}-\phi_{\rm bg}.
\label{baregeometric}
\end{equation}
Note the sign defining $\phi_{\rm bg}$, which is chosen to keep geometric phases predominantly positive in our discussion.

Spin dynamics is calculated in a similar manner using
the spin Hamiltonian $H_{\rm S}={\bf H(\theta)}\cdot {\bm \sigma}$, which gives
the spin rotation operator $\exp( iH_{\rm S} dt)$, where ${\bm \sigma}$ is the vector of Pauli matrices.
Here, ${\bf H}=(H_x,H_y)$ are parameters of the
Hamiltonian during a cycle.  The Hamiltonian features a degeneracy
point at ${\bf H}={\bf 0}$, which for $H_0>H_1$ moves outside the path sweeped by the parameters during a cycle,
marking a change in topology (Fig.~\ref{fieldtopo}). 
For an arbitrary normalized initial spin-$1\over 2$ spinor $\psi_i$, the final state after a roundtrip around the circuit { can be calculated by applying the propagator $n$ times: 
\begin{equation}
\psi_f=\prod_{m=1}^{n}\exp( i{\bf H(\theta_{m})}\cdot {\bm \sigma} d\theta) \psi_i ,
\label{matrix}
\end{equation}
where $\theta_{m}=(m-1)d\theta$ and $d\theta=2\pi/n$, with $n\rightarrow\infty$. We take $n$ finite but large (of the order of 5000) and check convergence of the numerical solution}. The total U(1) phase change $\phi_{\rm tot}$ is obtained directly from the phase acquired in the state evolution. Besides, the bare spin dynamic phase is calculated from
\begin{equation}
\phi_{\rm sd}=  \int_0^{2\pi} \langle \psi|H_{\rm S} |\psi \rangle d\theta.
\label{dynamicsu2}
\end{equation}
From this, the bare spin geometric phase is extracted from the difference between $\phi_{\rm sd}$ and $\phi_ {\rm tot}$ as in Eq.~(\ref{baregeometric}).



\section{Geometric phases for SO(3)}

\begin{figure*}
\includegraphics[width=\columnwidth]{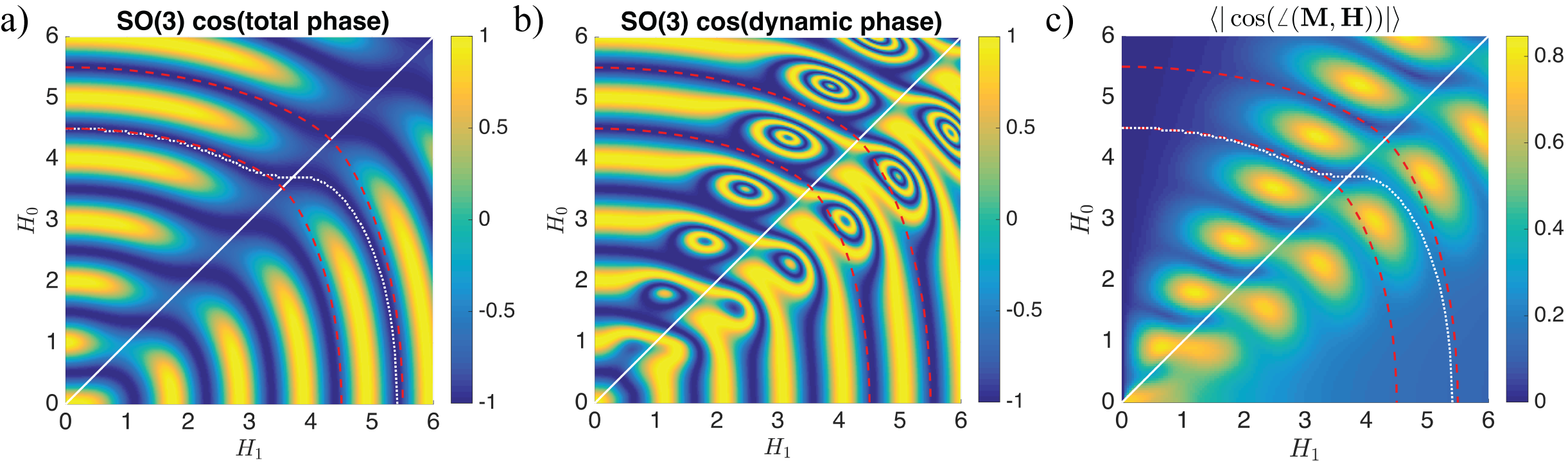}
\caption{The total phase (a) and the dynamic phase (b) in a single cycle of SO(3) rotations as a function of the $H_0$ and $H_1$ field components.
The total phase features a topological shift close to the critical line $H_0=H_1$ (solid line).
The dotted line gives the position of the 5th minimum where the total phase $\phi_{\rm tot}=4.5\times 2\pi$.
The lower and upper dashed red lines in (a) and (b) indicate where the adiabatic dynamic phase $\phi_{\rm ad}$ [Eq.~(\ref{wavefront})] is
$4.5\times 2\pi$ and $5.5\times 2\pi$, respectively.
c) Degree of nonadiabaticity quantified by $|\cos (\angle({\bf M},{\bf H}))|$ averaged over a cycle.
Values close to zero indicate approximately adiabatic evolution.
}
\label{fig-4s}
\end{figure*}
\begin{figure*}
\includegraphics[width=\columnwidth]{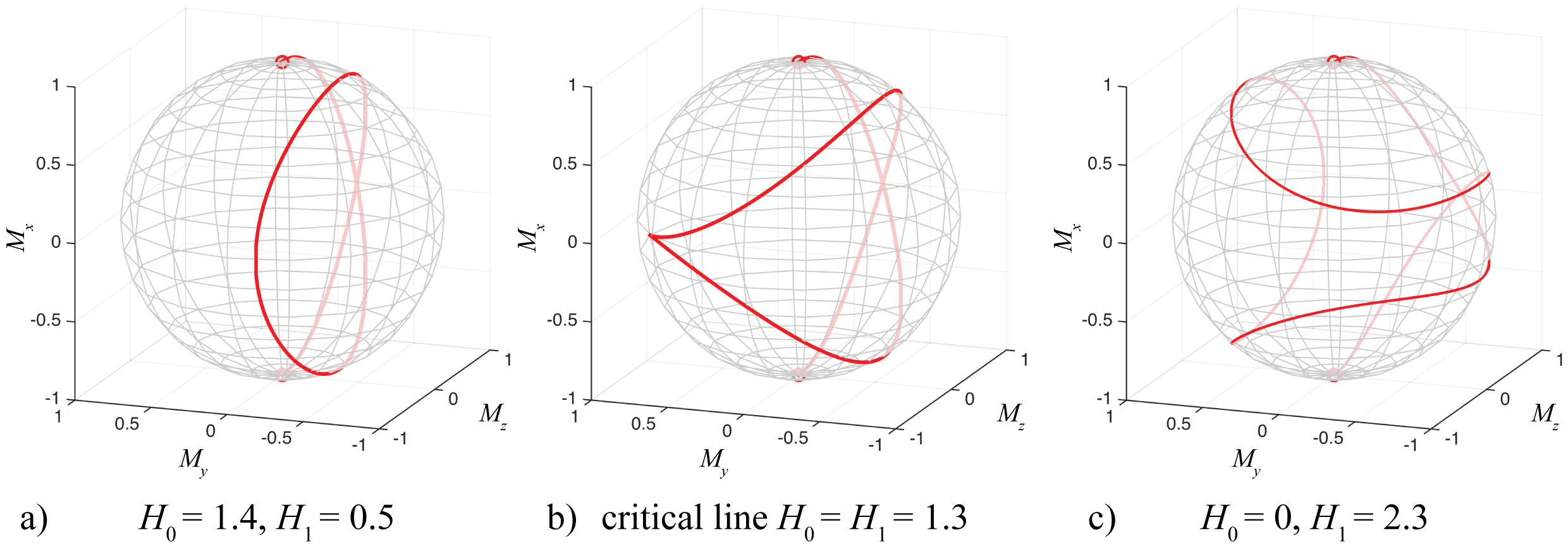}
\caption{Paths of the magnetic moment vector ${\bf M}$ calculated for the total phase $\phi_{\rm tot}=1.5\times 2\pi$ in different field topologies;
to the left of the critical line (a), at the critical line (b), and to the right of the critical line (c). The rotation angle of the magnetic moment vector (dynamical phase $\phi_{\rm d}$),
{\em i.e.} the path length from the 'north' pole to the 'south' pole in a given field, is $1.47\times 2\pi$ in (a), $1.68\times 2\pi$ in (b), and $2.20\times 2\pi$ in (c).
}
\label{fig-paths}
\end{figure*}

Geometric phases have been associated with adiabatic~\cite{berry} and nonadiabatic~\cite{aharonov} cyclic evolution.
Recently, the emergence of an \emph{effective} geometric phase has been reported in loop-shaped spin interferometers far from the adiabatic regime~\cite{saarikoskiprb}.
This effective geometric phase was found to be related to a topological transition in the quantum conductance occuring at $H_0=H_1$, and
it was explained in terms of the winding parity of the eigenstates around the poles of the Bloch sphere. In contrast to the smooth behavior displayed by
the effective geometric phase, the bare geometric and dynamic phases showed complex correlated patterns close to the topological transition.

Here we show that an analogous effective geometric phase can be also identified when SO(3) rotations of classical magnetic moments are considered.
We first study the evolution of a classical magnetic moment initially oriented along the $x$-axis, ${\bf M}_i={\bf e}_{x}$
(perpendicular to the driving field at $\theta=0$), as a result of applying the rotation matrix~(\ref{class_matrix}).
Due to precession, the magnetic moment acquires SO(3) phases which can be evaluated as detailed in the previous Section.
Cosines of the total phase $\phi_{\rm tot}$ and
the bare dynamic phase $\phi_{\rm d}$ in a roundtrip in the field  are shown in Figs.~\ref{fig-4s}-a and b.
When the rotating field component $H_1$ is absent, the magnetic moment precesses uniformly and the resulting total phase
is equal to the dynamic phase $\phi_{\rm tot}=\phi_{\rm d}=2\pi H_0$ according to Eqs.~(\ref{totphase}) and (\ref{dynamicso3}).
Precessions give rise to oscillations shown close to the vertical axis of Fig.~\ref{fig-4s}-a.
These are analogous to the Zeeman oscillations [U(1) phase rotation] in the case of spin transport in mesoscopic loops~\cite{saarikoskiprb}.
When both field components are finite the evolution of the magnetic moment is complex.
The dynamic phase $\phi_{\rm d}$ in Fig.~\ref{fig-4s}-b 
displays a rich texture due to nonadiabatic dynamics. The bare geometric phase $\phi_{\rm bg}=\phi_{\rm d}-\phi_{\rm tot}$,
not depicted here, shows a complementary complexity. Still, the total phase has a smooth behaviour and we note a distinct phase shift close to the critical line $H_0=H_1$
(see Fig.~\ref{fig-4s}-a). 
We associate this phenomenon with a change in the topology of the driving field and an emerging effective {\em classical} geometric phase in the following.

Figure~\ref{fig-paths} illustrates the magnetic moment dynamics at a fixed total phase $\phi_{\rm tot}={3\over 2}\times 2\pi$, 
corresponding to paths starting at the `north' pole of the sphere and ending in the `south' pole, for different field topologies.
Notice that lines of constant total phase\cite{footnote} 
equal to a half-integer multiple of $2\pi$ are easily identified in
the $H_1-H_0$ plane as those where $\cos(\phi_{\rm tot})=-1$ (see, {\em e.g.}, the dotted line in Fig.~\ref{fig-4s}-a). These lines stretch from the $H_0$ axis to the $H_1$ axis. { Along them,} $\phi_{\rm tot}$
can be directly calculated since $\phi_{\rm tot}=\phi_{\rm d}=2\pi H_0$ at the $H_0$ axis. In our case, when $H_0=1.5$ and $H_1=0$, the tip of
the magnetic moment describes a great circle with length $1.5 \times 2\pi$, equal to the dynamical phase at this point. As we approach the $H_1$ axis,
the total rotation angle increases and so does the dynamical phase, which is equal to  $\phi_{\rm d}=2.20 \times 2\pi$ at $H_0=0$, $H_1=2.32$ (Fig.~\ref{fig-paths}-c).
This angle $\phi_{\rm d}$ is slightly less than $H_1 \times 2\pi$ due to nonadiabatic effects. In view of Eq.~(\ref{baregeometric}),
it is evident that the increase of the total rotation angle is also associated with an increase in the bare geometric phase.

Close to the critical line $H_0=H_1$ the magnetic field is weak and it changes direction rapidly at $\theta=\pi$.
The magnetic moment cannot reorient fast enough causing it to lose the perpendicular
orientation with respect to the magnetic field.  This is manifested as an increase in nonadiabaticity, quantified here by $|\cos(\angle({\bf M},{\bf H}))|$ averaged over a cycle
of the field (Fig.~\ref{fig-4s}-c). Values close to zero indicate approximately adiabatic evolution.
Nonadiabatic { features are apparent} close to the critical line in Fig.~\ref{fig-4s}-c, causing the complex patterns in the dynamic phase in Fig.~\ref{fig-4s}-b.
However, we find that the evolution is approximately adiabatic when the total phase is a half-integer times $2\pi$, and this approximation improves with field strength,
see {\em e.g.} the dotted line corresponding to $\phi_{\rm tot}=4.5\times 2\pi$ in Fig.~\ref{fig-4s}-c. We use this fact to extract the effective geometric phase.

{First, we introduce the adiabatic dynamic phase in a closed path,
$\phi_{\rm ad}$, as the phase acquired in an homogeneous field of
strength $H_{\rm ave}=\int_0^{2\pi}d\theta\, |{\bf H}|/2\pi$, which is
the average field in a roundtrip. An explicit calculation gives
\begin{eqnarray}
\phi_{\rm ad} &=\sqrt{H_0^2+H_1^2} \int_0^{2\pi}d\theta\, \sqrt{1+{2H_0H_1 /(H_0^2+H_1^2) }\cos\theta} \nonumber \\
 &= 2  (H_{\rm 1} + H_0)\left [( E(\pi/4, {\cal H}  ) + E(3\pi/4, {\cal H} ) \right ],
\label{wavefront}
\end{eqnarray}
where} ${\cal H} = 4H_{1} H_0/(H_{1}+H_0)^2$, and $E(\varphi,m)$ are elliptic integrals of the 2nd kind.
Notice that $\phi_{\rm ad}$ may be understood as a sort of a smooth component of the bare dynamic phase $\phi_{\rm d}$.
Fig.~\ref{fig-4s}-a clearly shows quasi-adiabatic characteristics in the form of wavefronts at constant adiabatic dynamic phase (dashed lines).

\begin{figure*}
\includegraphics[width=\columnwidth]{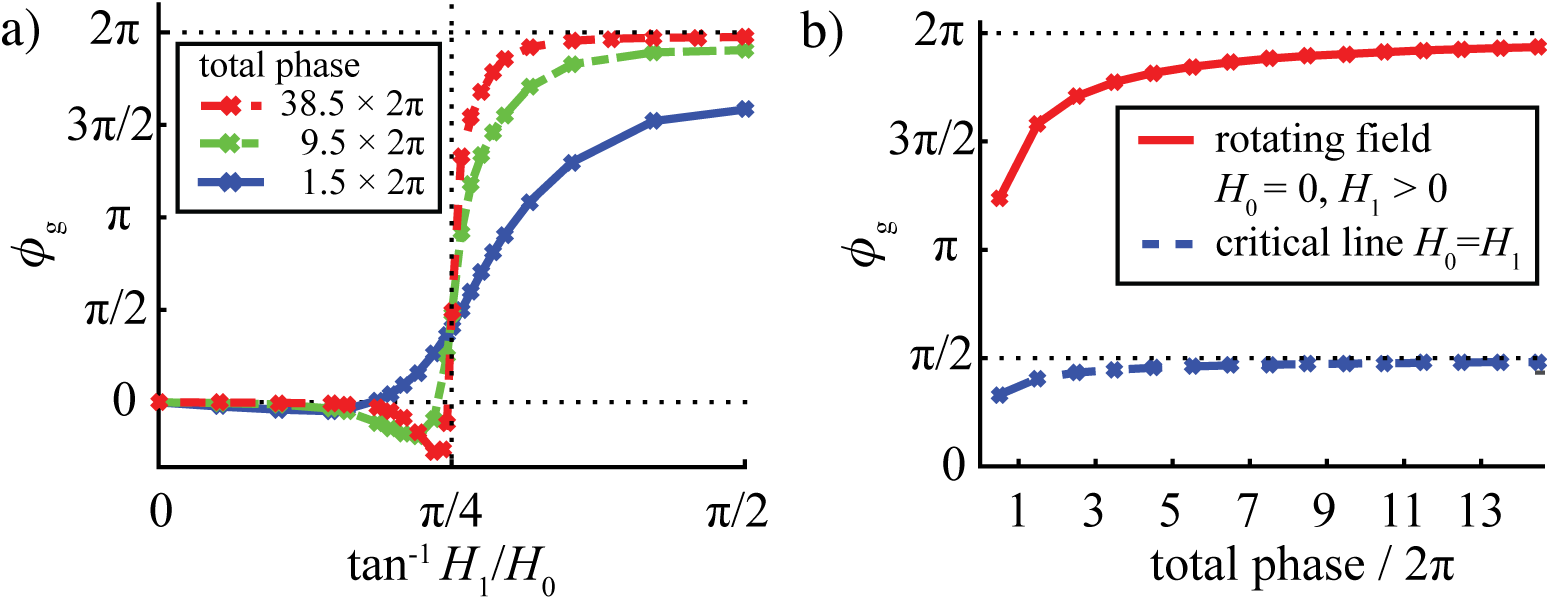}
\caption{a) The effective geometric phase $\phi_{\rm g}$ for SO(3) rotations calculated
in the adiabatic approximation [Eq. (\ref{effectivegeometric})]
as a function of field component angle $\tan ^{-1} H_1/H_0$. The critical line is the vertical dotted line at $\pi/4$.
In the limit of high fields the phase shift approaches a step-like pattern.
b) $\phi_{\rm g}$ as a function of the total phase shown for a rotating field component only ($H_0=0$ and $H_1>0$,
the horizontal axis in Fig.~\ref{fig-4s}-a), and at the critical line ($H_0=H_1$, the diagonal line in Fig.~\ref{fig-4s}-a).
In the limit of strong fields $\phi_{\rm g}$ approaches $2\pi$ and ${\pi \over 2}$, respectively, corresponding to
the adiabatic Berry phases.}
\label{fig-2s}
\end{figure*}

We now define the effective geometric phase as the difference between the adiabatic dynamic phase and the total phase
\begin{equation}
\phi_{\rm g} = \phi_{\rm ad}-\phi_{\rm tot}.
\label{effectivegeometric}
\end{equation}
Figure~\ref{fig-2s}-a shows $\phi_{\rm g}$ as a function of the angle $\tan^{-1} H_1/H_0$ in the $H_1-H_0$ plane.
Two limits are clearly distinguished: $\phi_{\rm g}\rightarrow 0$ for $H_0 \gg H_1$ while $\phi_{\rm g}\rightarrow 2\pi$ when $H_1 \gg H_0$.
These limits correspond to the classical SO(3) adiabatic Berry phases for the two different field topologies shown in Fig.~\ref{fieldtopo}.
It is worth noting that, despite the abrupt change in the field topology, the effective geometric phase undergoes
a smooth transition that nevertheless features a step-like behaviour in the limit of high fields.
Similar results have been found in the case of spin magnetic moments for loop-shaped interferometers~\cite{saarikoskiprb}.
This effect is therefore an SO(3) counterpart of the Lyanda-Geller's topological transition of
the geometric phase for spin rotations~\cite{lyanda-geller}, but involves an effective geometric phase instead of the bare geometric phase,
and displays a smooth transition instead of an abrupt change~\cite{lyanda-geller}.

The evolution of the classical magnetic moment is in general nonadiabatic close to the critical line.
However, exactly at the critical line $H_0=H_1$ the total field changes slowly and reverses its direction at $\theta=\pi$.
This causes a cusp in the path of the magnetic moment in Fig.~\ref{fig-paths}-b. Adiabaticity is then enhanced and an effective geometric phase
$\phi_{\rm g}={\pi/2}$ is obtained in the limit of high fields~\cite{footnote-2} (see Fig.~\ref{fig-2s}-b).

The bare geometric phase in the classical SO(3) case is related to the geometry of the path described by the tip of the classical magnetic moment
on the sphere ${\rm S}^2$. 
This is also approximately true for the {\em effective} geometric phase along the
lines where $\phi_{\rm tot}$ is a half-integer due to nearly  adiabatic evolution 
[{\em i.e.} using $\phi_{\rm d}\approx \phi_{\rm ad}$ and Eq. (\ref{effectivegeometric})].
Therefore, at fixed half-integer total phase, the increase in the length of this path (the total rotation angle) 
is associated with a change in $\phi_{\rm g}$ (see the paths in Fig.~\ref{fig-paths}-a--c and the calculated effective geometric phases in Fig.~\ref{fig-2s}-a).
The smooth shift in the cosine of the total phase close to the critical line (Fig.~\ref{fig-4s}-a) suggests that this result applies more generally to other than half-integer total phases.
Again, this phenomenon is analogous to the case discussed in the context of spin transport in Ref.~\cite{saarikoskiprb},
where the effective geometric SU(2) phase was found to be connected to the rotation of the spin in the Bloch sphere.

\section{SU(2) spin rotation}

\begin{figure}
\includegraphics[width=0.5\columnwidth]{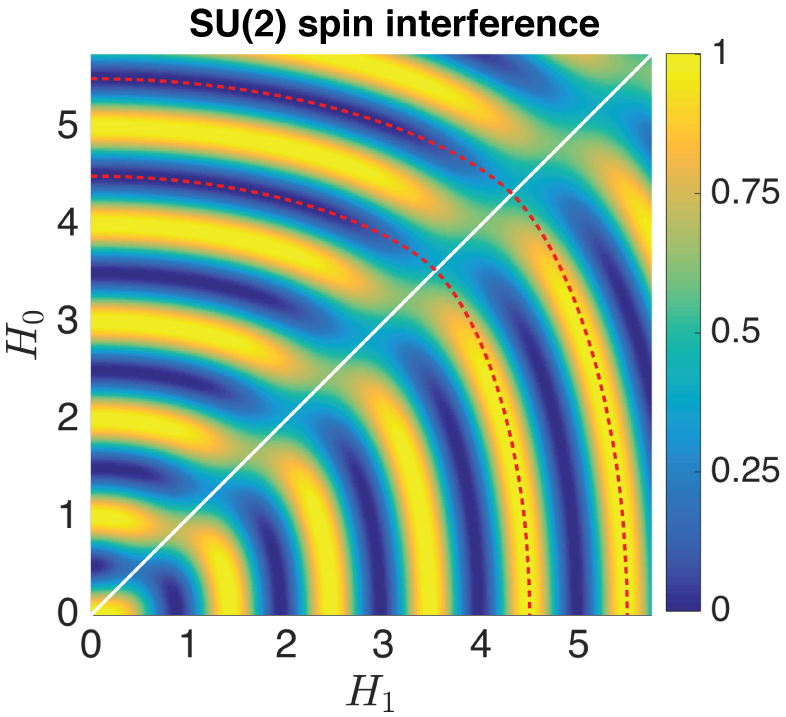}
\caption{Topological transition in the spin interference pattern for SU(2) rotations of a spin as
a function of $H_0$ and $H_1$ field components. The transition features a shift in the interference peak positions close to the critical line $H_0=H_1$ (solid line).
The lower and upper dashed red lines indicate where the adiabatic dynamic phase $\phi_{\rm ad}$ [Eq.~(\ref{wavefront})] is
$4.5\times 2\pi$ and $5.5\times 2\pi$, respectively.}
\label{fig-7}
\end{figure}
The effective geometric phase for SU(2) spin rotations was identified in loop-shaped spin interferometers subject
to an effective magnetic field of the form given in Eq.~(\ref{magfield}) (see Ref.~\cite{saarikoskiprb}).
In this last work, {  the $H_1$ component of the magnetic field in the spin-rotation model of Eq. (\ref{matrix}) was associated with the Rashba spin-orbit interaction strength in semiconductor quantum wells}.
Here we examine the symmetries of the interference pattern in this spin-$1\over 2$ system and compare it with the classical SO(3) case.

Figure~\ref{fig-7} shows the interference pattern for SU(2) spin precessions in the loop configuration. More precisely, we calculate the interference of waves from 
 {
\begin{equation}
{1\over 2}|\psi_i+\psi_f|^2=1+{1\over 2}(\psi_i\psi_f^\dag+\psi_f\psi_i^\dag),
\label{spininterference}
\end{equation}
where $\psi_i$ denotes an arbitrary normalized} initial spin orientation and $\psi_f$
is the spinor after one cycle around the loop, calculated using the method detailed in Ref.~\cite{saarikoskiprb} together with Eq.~(\ref{matrix}).
This calculation corresponds to a weakly coupled loop where multiple windings around the loop can be neglected~\cite{footnote-3}.
We obtain results recalling adiabatic spin dynamics in a nonadiabatic scenario, giving rise
to wavefronts at approximately constant adiabatic dynamic phase [Eq.~(\ref{wavefront})].
At the critical line, the total probability amplitude remains finite due to complex nonadiabatic spin dynamics and a topological transition is visible.
We find a close similarity in how the topological transitions emerge for both SU(2) spin rotations and classical SO(3) rotations (see Fig~\ref{fig-4s}-a).

\label{sec:interference}

A detailed analysis of the results reveals that the interference pattern { obtained from Eq.~(\ref{spininterference}), depicted in Fig.~\ref{fig-7},} is independent of the initial spin orientation and phase
[${\rm SU(2)}\times {\rm U(1)}$ symmetry], which contrasts with the classical SO(3) case where the total phase producing
the corresponding interference pattern (Fig.~\ref{fig-4s}-a) depends on the initial magnetic moment orientation.
This is due to the existence of an additional ${\rm U(1)}$ symmetry only present in the quantum mechanical case.
Despite this, we find that decomposition of the total quantum phase in the bare dynamic and geometric components actually depends on $\psi_i$.
As an example, consider the oscillations along the $H_{0}$ axis in Fig.~\ref{fig-7} and the associated phases.
Here, an initial spin orientation parallel to the field results in a total phase with no geometrical contribution, which however is
present for an initial perpendicular orientation due to precessions.

\section{Periodic states}

\begin{figure*}
\includegraphics[width=\columnwidth]{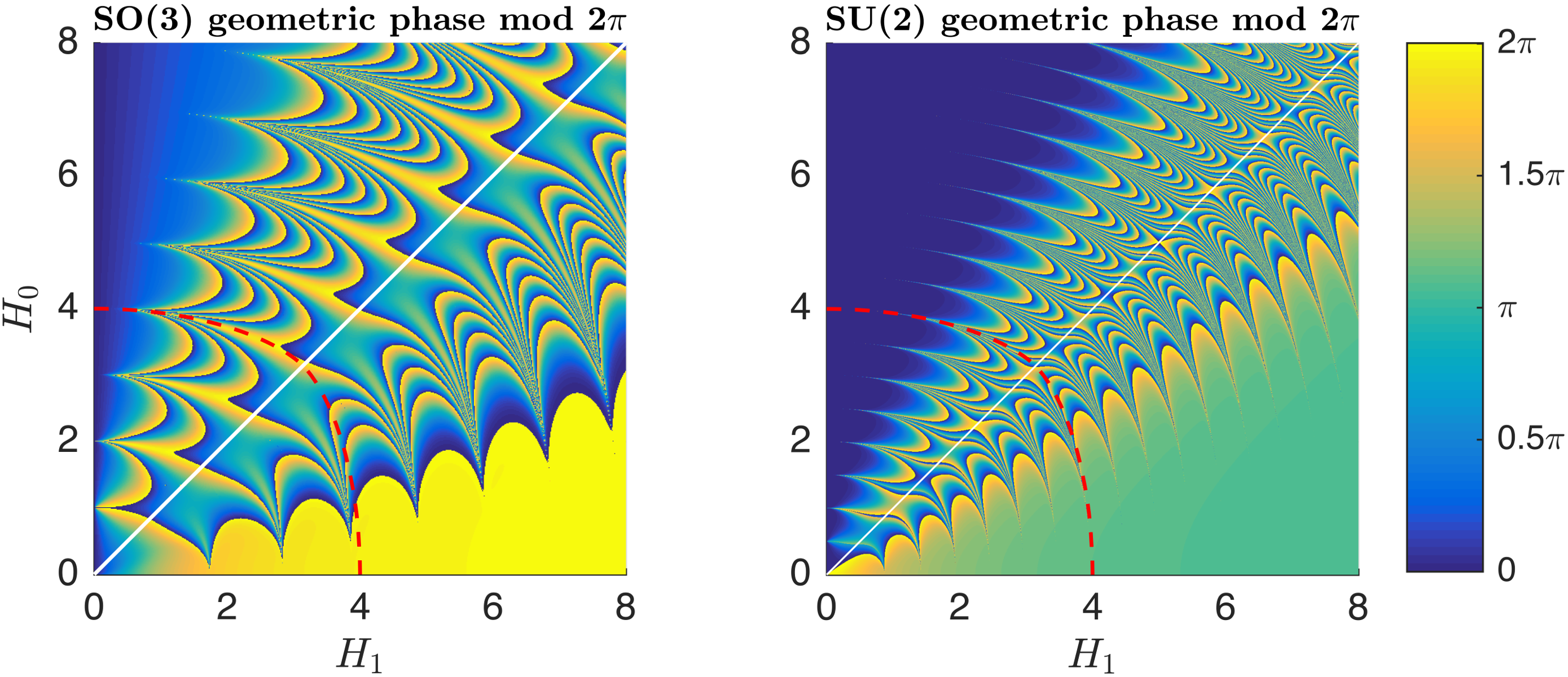}
\caption{The bare geometric phase $\phi_{\rm bg}$ modulo $2\pi$ displays a topological
phase shift across the critical line (solid lines) both for SO(3) (left)
and SU(2) (right) eigenstates
The geometric phase at $H_1=0$ vanishes and the values in the limit of high $H_1$ and $H_0\ll H_1$ are $2\pi$ and $\pi$ for SO(3) and SU(2), respectively,
matching the adiabatic Berry phases.
Close to the critical line precessions give a large bare geometric phase leading to the complex patterns in the figures.
The adiabatic dynamic phase $\phi_{\rm ad}=4\times 2\pi$ at the dashed lines.}
\label{fig-geo}
\end{figure*}
\begin{figure*}
\includegraphics[width=\columnwidth]{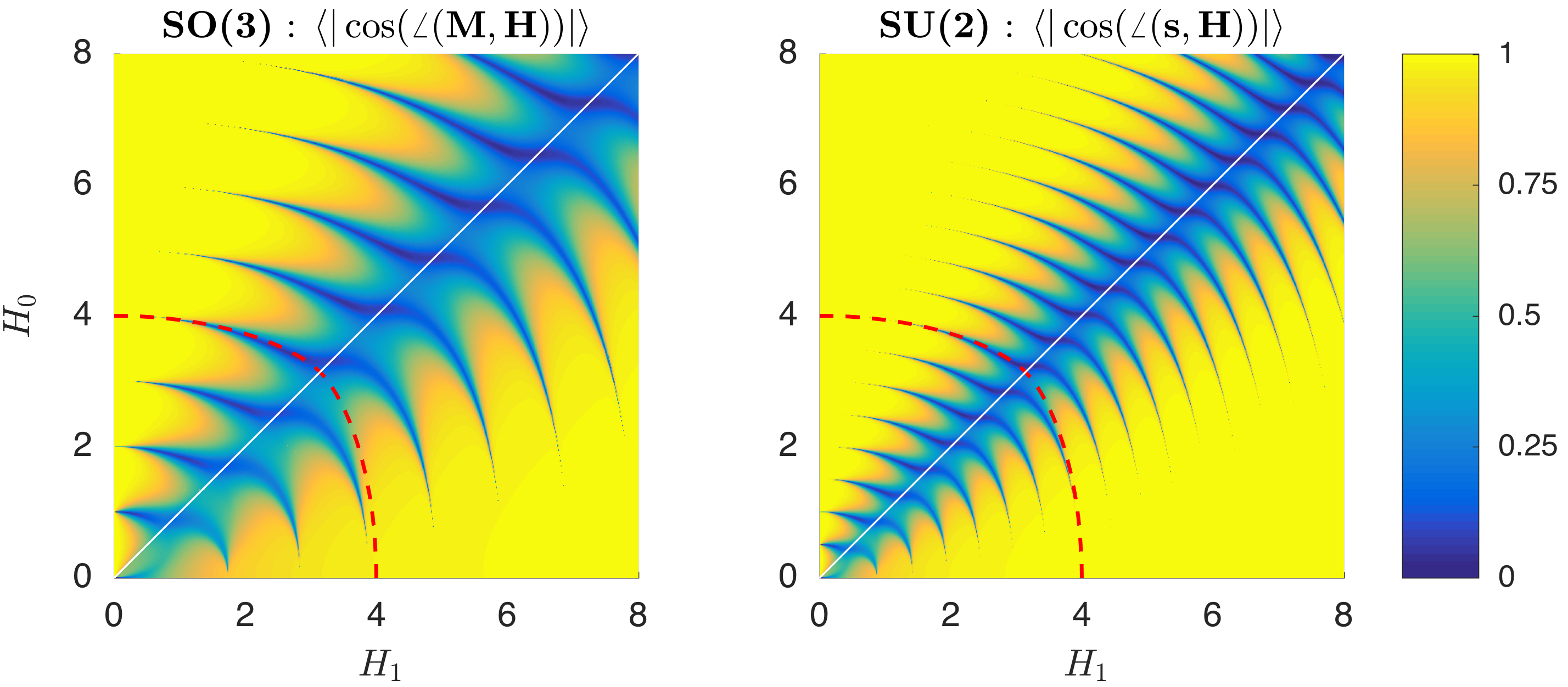}
\caption{Averaged absolute value of the projection of the driving field to the corresponding magnetic moment $\bf M$ eigenstates (left) and
spin ${\bf s}$ eigenstates (right).
The field and magnetic moment (spin) orientations are approximately parallel for $H_0\gg H_1\gg 1$ and $H_1\gg H_0\gg 1$.
Close to the critical line the angle is close to $90^\circ$ and dynamics are dominated by precessions.
The adiabatic dynamic phase $\phi_{\rm ad}=4\times 2\pi$ at the dashed lines.}
\label{fig-perp}
\end{figure*}

\label{eigenstates}

{ In the classical case,} differences between effective and bare geometric phases become apparent when we consider the evolution of eigenstates after
a full cycle of the total magnetic field (Floquet states). We use the fact that the total phase change in a full cycle of SO(3) vanishes for eigenstates
(up to integer multiple of $2\pi$) which, using Eq. (\ref{baregeometric}), means that bare dynamic and geometric phases are equal (modulo $2\pi$).
In the following, we study eigenstates of SO(3) rotations and show a topology-related shift in the calculated bare geometric phase associated with the effective geometric phase.

The classical eigenstates describe periodic states, and they are given by the
solutions of the equation $R{\bf M}(0)={\bf M}(0)$, where $R$ is the matrix governing the classical rotations
defined in Eq.~(\ref{class_matrix}). The bare geometric phase $\phi_{\rm bg}$ (modulo $2\pi$) for SO(3) eigenstates is
shown in the left panel of Fig.~\ref{fig-geo}. It features a complex structure close to the critical line which contrast with the smooth behavior outside this region.
The wide regions of nearly constant bare geometric phase close to the $H_0$ and $H_1$ axis
resemble a topological phase shift related to the 
emerging effective geometric phase. In particular, close to the
horizontal axis the effective geometric phase coincides with
the adiabatic Berry phase, {\em i.e.} $2\pi$.
The bare geometric phase for SU(2) spin eigenstates, obtained from Eq.~(\ref{matrix}), displays an analogous structure
(right panel of Fig.~\ref{fig-geo}; see also Ref.~\cite{saarikoskiprb}), with the adiabatic Berry phase equal to $\pi$ in this case.


Left panel of Fig.~\ref{fig-perp} shows $|\cos(\angle({\bf M},{\bf H}))|$ averaged over a cycle.
These results demonstrate that close to the critical line the eigenstate is almost perpendicular to the driving field.
This favours precessions, which give rise to large dynamic and geometric SO(3) phases resulting in the complex pattern shown in the left panel of Fig.~\ref{fig-geo}.
In contrast, far from this region the eigenstates are approximately parallel to the driving field.
We find analogous dynamics for SU(2) spin eigenstates (right panel of Fig.~\ref{fig-perp}),
displaying a double frequency pattern due to spin evolution on the Bloch sphere.
Spin precessions close to the critical line give a large bare geometric phase
which is the dominating contribution to the total phase.


\section{Discussion and outlook}

We have shown that an effective geometric phase emerges in classical systems  where the topology of the driving field changes.
The effective geometric phase is analogous to that found in quantum spin-${1\over 2}$ systems.
{ These results are therefore very general. In addition, they are linked to the properties of the SO(3) group of rotations~\cite{milman}.
The difference between the values obtained for the effective geometric phase in  the adiabatic limit 
 ($2\pi$ and $\pi$ in the classical and quantum cases, respectively) is related to the fact that SO(3) is doubly connected
while SU(2) is simply connected.
Despite the abrupt change in the field topology, the effective geometric phase 
shows a smooth but distinct transition close to the critical line of the phase boundary due to nonadiabatic effects,
both in the quantum as well as in the classical case.}

Applications of the theory would require a method to measure the resulting phases, {\em e.g.}, via interference,
resonance or a direct measurement of the magnetic moment orientation.
In the case of spin rotations applications may include electron spin phases in mesoscopic spin systems or nuclear magnetic resonance experiments.
Topological transitions of the effective geometric phases could be probed
also in spins other than $1 \over2$ such as
bosons~\cite{tomita}, spin-$3 \over 2$ of holes~\cite{arovas} or nuclei in nuclear quadrupole resonance spectroscopy (NQR)~\cite{das,tycko}.
Classical systems could include macrospins in ferromagnetic resonance experiments\cite{chikazumi,gurevich} if sufficiently high rotating fields
can be generated.

We acknowledge fruitful discussions with Yasunobu Nakamura, Gerrit E. W. Bauer, and Fumiya Nagasawa.
This work was supported by  Japan Society for the Promotion of Science with Grant-in-Aid for Scientific Research C
No. 26390014 (H. S.) and Specially Promoted Research No. 15H05699 (J. N.).
D. F. and J. P. B. acknowledge support from project No. FIS2014-53385-P (MINECO, Spain) with FEDER funds.


\begin{thebibliography}{99}
\bibitem{footnote-2} This is incidentally the same value as for SU(2) spin rotations at the critical line~\cite{lyanda-geller}.
However, SU(2) spin dynamics at the critical line displays a $4\pi$ periodicity in $\theta$ (Ref.~\cite{aharonov}),
and the reversal of the field direction at $\theta=\pi$ is associated with a band transition.
\bibitem{footnote-3} In Ref.~\cite{saarikoskiprb}, only periodic spin eigenstates were considered. Here, on the contrary, we present results corresponding to
the interference of an arbitrary spinor before and after coming across the loop.
\end{thebibliography}
\end{document}